\documentclass[twocolumn,preprintnumbers,showpacs,prb,a4paper]{revtex4}

\usepackage{graphicx}
\usepackage{color}
\usepackage{amsmath}
\usepackage{amssymb}
\usepackage{subfigure}
\usepackage{epsfig}

\newcommand{\ket}[1]{\ensuremath{|#1\rangle}}
\newcommand{\bra}[1]{\ensuremath{\langle#1|}}
\newcommand{\dmin}{\ensuremath{D^-}}
\newcommand{\dplus}{\ensuremath{D^+}}
\newcommand{\dzero}{\ensuremath{D^0}}

\begin{document}
\title{Optically induced spin to charge transduction in donor spin read-out.}

\author{M. J. Testolin}
\author{Andrew D. Greentree}
\author{C. J. Wellard}
\author{L. C. L. Hollenberg}
\affiliation{Centre for Quantum Computer Technology, School of Physics, University of Melbourne, VIC 3010, Australia}

\begin{abstract}
The proposed read-out configuration \dplus\dmin\ for the Kane Si:P architecture[Nature {\bf 393}, 133 (1998)] depends on spin-dependent electron tunneling between donors, induced adiabatically by surface gates.  However, previous work has shown that since the doubly occupied donor state is so shallow the dwell-time of the read-out state is less than the required time for measurement using a single electron transistor (SET).  We propose and analyse single-spin read-out using optically induced spin to charge transduction, and show that the top gate biases, required for qubit selection, are significantly less than those demanded by the Kane scheme, thereby increasing the \dplus\dmin\ lifetime.  Implications for singlet-triplet discrimination for electron spin qubits are also discussed. 
\end{abstract}


\maketitle

\section*{Introduction}
Solid-state quantum computer (SSQC) architectures are of particular interest for the development of a working quantum computer, as any such architecture could leverage the power of the semiconductor industry for scalability.  The Kane architecture\cite{Kane} is one contender for an SSQC.  Here the qubits are phosphorus donors in isotopically pure $^{28}$Si with $I=0$.  The logical state of the qubit is encoded on the nuclear spin of the phosphorus donor which has nuclear spin $I=1/2$.  The advantage of encoding the qubit in this way is that these Si:P systems are known to exhibit long relaxation times\cite{Feher,Wilson}, meaning the nuclear spin is highly robust to decoherence.  On the other hand, weak coupling to the environment (and hence a measurement device) renders measurement of the spin qubit extremely difficult.  All operations are dependent on electron mediated interactions with the nucleus via the hyperfine interaction.

Donor electron spin based proposals for an SSQC\cite{Vrijen,Hill,deSousa} are also of interest.  Electron spin qubits may offer enhanced simplicity for qubit control, read-out and gate operation speed (for exchange based proposals) over their nuclear spin counterpart.  Recent measurements\cite{Tyryshkin} of the electron spin coherence time, $T_2$, for phosphorus donors in Si, give $T_2>60\ \mbox{ms}$ at $7\ \mbox{K}$.  Despite the electron spin coherence time being shorter than the coherence time for nuclear spins, relatively faster gate operations mean that of order $10^6$ operations are possible within the coherence time\cite{Hill}.

Measurement and intialisation are essential requirements of quantum computation.  Experimental detection of a single electron spin in solid-state systems has only recently been reported.  Detections of a single electron spin have now been made in a quantum dot system formed in the two-dimensional electron gas (2DEG) of a GaAs/AlGaAs heterostructure\cite{Elzerman}, via magnetic resonance force microscopy\cite{Rugar} in SiO$_2$ and optically in nitrogen-vacancy (NV) defect centres in diamond\cite{Jelezko}.  Proposals exist for single-spin read-out within a number of different qubit systems, ranging from spin to charge transduction techniques involving:  adiabatic transfer\cite{Kane}, spin valves\cite{Loss}, gated resonant transfer\cite{Hollenberg}, asymmetric confining potentials\cite{Friesen} and spin-dependent charge fluctuations\cite{Barrett}.  Other novel methods include ancilla assisted read-out\cite{Greentree,Ionicioiu} and optical read-out\cite{Fu,Shabaev}.  In all spin to charge transduction processes, measuring the state of the nuclear spin qubit is turned into the task of measuring a spin-dependent electron charge transfer event: for example in Kane, the process whereby a two neutral donor system, \dzero\dzero\ becomes \dplus\dmin.  Indirect measurement of the spin state of the qubit in this way is possible due to the relative ease of coupling to a charge measurement device, e.g. a single electron transistor (SET) or quantum point contact (QPC).

The resultant doubly occupied state, \dmin, is very shallow, with a binding energy of $\sim 1.7\ \mbox{meV}$\cite{Taniguchi,Norton} and hence may be easily ionised.  Read-out via the adiabatic Kane protocol requires electric fields which may be too large to preserve the \dmin\ state long enough for detection by the SET.  In particular, the maximum DC field strength tolerated for a ``safe'' \dmin\ dwell time of $T_{\dmin}\approx 10\ \mu\mbox{s}$ has been estimated to be an order of magnitude smaller than the field required for adiabatic charge transfer (see Sect.~\ref{Sect:GRST})\cite{Hollenberg}.

We describe a means by which to perform the resonant spin-dependent charge transfer proposed in Ref.~\onlinecite{Hollenberg} utilising a far infrared (FIR) laser at resonance with the transition \dzero\dzero$\longrightarrow$\dplus\dmin. This FIR laser induced resonant transfer is related to that implied by Larionov \emph{et~al.}\cite{Larionov} in generating qubit gates in a \dmin\ based quantum computer proposal. In an electron spin architecture, optically driven spin to charge transduction would be a means by which to perform singlet-triplet read-out, which is sufficient for cluster state quantum computation\cite{Rudolph}.

\section{Gated Resonant Spin Transfer}
\label{Sect:GRST}
The short lifetime of the \dmin\ state in the presence of a DC electric field motivates the proposal for gated resonant spin transfer\cite{Hollenberg}.  The adiabatic charge transfer proposed by Kane relies on a slowly varying DC field to effect the \dzero\dzero$\longrightarrow$\dplus\dmin\ transition.  The shallow phosphorus donors are $45.5\ \mbox{meV}$ below the conduction band edge and the doubly occupied \dmin\ state is bound by only 1.7 meV.  The problem with the existing adiabatic charge transfer scheme is that application of the static DC field is likely to ionise the \dmin\ state to the conduction band.  By using additional suitably placed gates it may be possible to protect the system from ionisation during the read-out process, however, this would require the fabrication and control of complex arrays of gate structures.  The measurement time for small induced charge levels ($\Delta q<0.05q_e$) using single-shot read-out with a radio-frequency single electron transistor (rf-SET) operating near the quantum limit is of the order $T_{\rm SET}\approx 1\ \mu\mbox{s}$\cite{Buehler}.  This means that for read-out to be successful, the survival of the \dmin\ state must be longer than $T_{\rm SET}\approx 1\ \mu\mbox{s}$.

In order to quantify this, the maximum DC field strength $F_{\rm DC}^*$ for a ``safe'' \dmin\ dwell time of  $T_{\dmin}\approx 10\ \mu\mbox{s}$ was calculated in Ref.~\onlinecite{Hollenberg}.  The DC field ($F_{\rm DC}^{ad}$) required in order to adiabatically transfer the charge between the two donors was also calculated as in the earlier work of Fang \emph{et~al.}\cite{Fang} and found to be much greater than $F_{\rm DC}^*$ for all cases of donor separation tested.  Specifically, for a donor separation of $R=30\ \mbox{nm}$, $F_{\rm DC}^{ad}/F_{\rm DC}^*\approx 11$.  Essentially this means that the read-out proposal based on Kane adiabatic charge transfer is problematic as the \dmin\ state is not sufficiently long-lived for high-fidelity SET detection.

\begin{figure}[tb]
\includegraphics{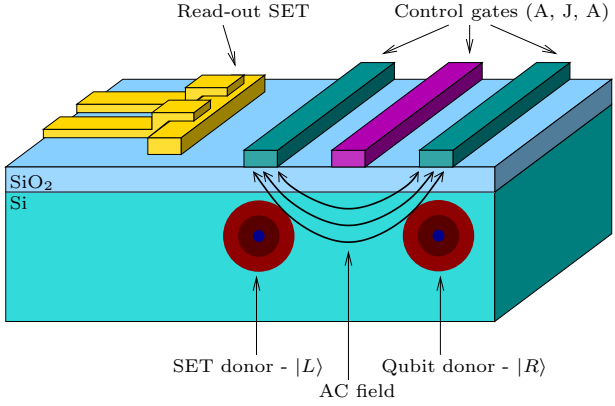}
\caption{(Colour online) Schematic of the device for the resonant spin dependent charge transfer of a single electron.}
\label{fig:resdevice}
\end{figure}

Gated resonant spin transfer was proposed in Ref.~\onlinecite{Hollenberg} as an alternative to the adiabatic charge transfer scheme of Ref.~\onlinecite{Kane}.  The idea behind gated resonant spin transfer is to replace the adiabatic DC electric field ($F_{\rm DC}^{ad}$) with a small DC electric field $F_{\rm DC}\ll F_{\rm DC}^*$ for qubit selection and an AC electric field with amplitude $F_{\rm AC}\ll F_{\rm DC}^*$ at resonance with the energy gap $\Delta E(F_{\rm DC})$ of the two states, \dzero\ and \dmin.  A schematic of the device can be seen in Fig.~\ref{fig:resdevice}. 

We begin by studying the dynamics of the \dzero\dzero$\longrightarrow$\dplus\dmin transition driven by gate fields only.  The Hamiltonian for the system is
\begin{equation}
\label{reshamil}
{\cal H}(t)={\cal H}_{0}+{\cal H}_{\rm DC}(t)+{\cal H}_{\rm AC}(t),
\end{equation}
where 
\begin{equation}
{\cal H}_{0} = -\frac{\hbar^2}{2m^*}\nabla_{1}^{2}-\frac{\hbar^2}{2m^*}\nabla_{2}^{2}-k\frac{q^2_e}{r_1}-k\frac{q^2_e}{r_2}-k\frac{q^2_e}{r_1^\prime}-k\frac{q^2_e}{r_2^\prime}+k\frac{q^2_e}{r_{12}}\nonumber
\end{equation}
and the DC and AC terms (applied along the donor separation axis, defined to be the x-direction) are given by
\begin{eqnarray}
{\cal H}_{\rm DC}(t) &=& q_e(x_1+x_2-R)F_{\rm DC}(t),\nonumber\\
{\cal H}_{\rm AC}(t) &=& q_e(x_1+x_2-R)F_{\rm AC}(t)\sin\omega t.\nonumber
\end{eqnarray}
${\cal H}_0$ is the ungated two-donor Hamiltonian in the effective mass approximation relevant for the Si system ($m^*=0.2m_e$, $\varepsilon=11.9\varepsilon_0$, $a_B=2\ \mbox{nm}$).  Here, $r_i^\prime=|\vec r_i-\vec R|$, where the $r_i$ give the electron positions relative to a phosphorus donor at the origin and $\vec R$ specifies the double donor separation.  The Coulombic constant relevant for the Si system is $k=1/4\pi\varepsilon$.  The $F_{\rm DC}(t)$ and $F_{\rm AC}(t)$ are square-pulses with time-dependence to signify that the turn-on times of these pulses will in general differ from each other.  Energies are scaled to the \dzero\ centre ground state energy in Si ($45.5\ \mbox{meV}$).

In this work we assume that the electric fields used to generate gated resonant spin transfer are uniform as described by ${\cal H}_{\rm DC}(t)$ and ${\cal H}_{\rm AC}(t)$.  A more complete analysis of the problem for specific gate structures would account for the non-uniformity of these fields, the effects due to mirror charges in the gates and the presence of charge traps at the SiO$_2$/Si interface.  The inclusion of a non-uniform field would alter the details of time scale and bias required for charge transfer, yet should not be too different from the analysis carried out here.  Including these effects is beyond the scope of this paper but presents an opportunity for future work.

At an operating temperature of $100\ \mbox{mK}$ the electrons will only occupy the $1s$ orbitals.  The starting state describes the \dzero\dzero\ system at $B=0$ with wave function 
\begin{equation}
\label{phiLR}
\psi_{LR}=N_{LR}(e^{-r_1-r_2^\prime}+e^{-r_1^\prime-r_2}).
\end{equation}
In this notation, $L$ and $R$ refer to the position of the electrons with respect to the left and right donors of a two donor system. The \dplus\dmin\ system is well described by the Chandrasekhar wave function
\begin{eqnarray}
\label{phiLL}
\psi_{LL}&=&N_{LL}(e^{-\alpha r_1 -\beta r_2}+e^{-\beta r_1 -\alpha r_2})(1+\lambda r_{12}),\\
\label{phiRR}
\psi_{RR}&=&N_{RR}(e^{-\alpha r_1^\prime -\beta r_2^\prime}+e^{-\beta r_1^\prime -\alpha r_2^\prime})(1+\lambda r_{12}).
\end{eqnarray}
The $\alpha$, $\beta$ and $\lambda$ are evaluated variationally\cite{Larsen}.  All total wave functions ($\Psi_{LL}$, $\Psi_{LR}$, $\Psi_{RR}$) are correctly anti-symmetrised when the spin component is considered, $\chi=\chi_{as}=[\ket{\negthickspace\downarrow\uparrow}-\ket{\negthickspace\uparrow\downarrow}]/\sqrt{2}$.  This spin singlet state is required for the charge transfer stage of read-out.

To effect a transition from \ket{LR} to the arbitrarily chosen doubly occupied \ket{LL} state, the gated fields described by ${\cal H}_{\rm DC}(t)$ and ${\cal H}_{\rm AC}(t)$ are pulsed and the transition from \ket{LR}$\rightarrow$\ket{LL} is studied numerically.

\begin{figure}[tb]
\centerline{\includegraphics{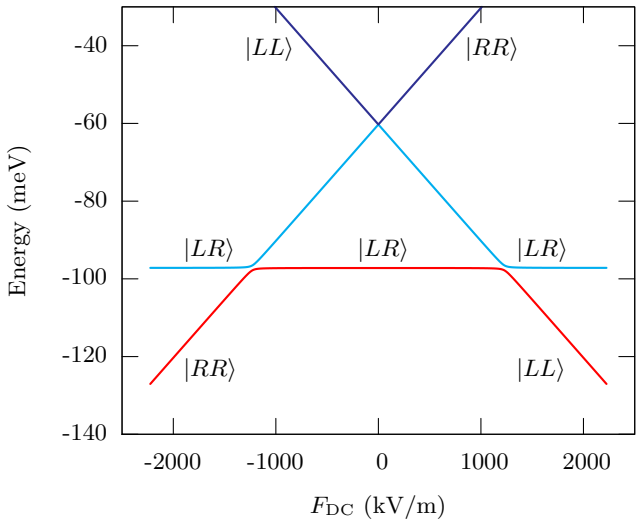}}
\caption{(Colour online) Energy level diagram as a function of DC electric field strength, $F_{\rm DC}$ for a donor separation of $R=30\ \mbox{nm}$.}
\label{fig:elevels30nm}
\end{figure}

\begin{figure}[tb]
\centerline{\includegraphics{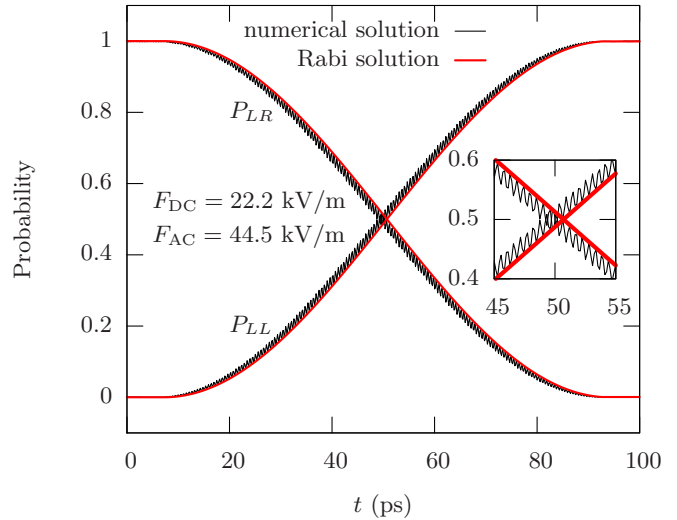}}
\caption{(Colour Online) Ground and excited state populations as a function of time showing spin-dependent charge transfer.  In the presence of a small DC electric field, $F_{\rm DC}$, the Rabi solution closely matches the numerical solution obtained from the three-state calculation.}
\label{fig:rabitrans}
\end{figure}

For the parameters considered, we find the lowest two states are effectively decoupled from the third state (see Fig.~\ref{fig:elevels30nm}), and we can therefore treat the system as an effective two-state system.
The Rabi solution\cite{Knight} to this problem gives the excited state population $P_{LL}$ (with ground state population $P_{LR}=1-P_{LL}$)
\begin{equation}
\label{rabi}
P_{LL}=\frac{|V|^2}{\hbar^2\Delta^2+|V|^2}\sin^{2}\Bigg(\sqrt{\Delta^2+\frac{|V|^2}{\hbar^2}}\frac{t}{2}\Bigg),
\end{equation}
where the dipole matrix element is 
\begin{equation}
V=q_e\bra{LL}(x_{1}+x_{2}- R)\ket{LR}F_{\rm AC}.
\end{equation}
$\Delta=\omega-\omega_0$ is the detuning of the AC field (with frequency $\omega$) with respect to the transition (with frequency $\omega_0$) and $R$ is the donor separation.  For the case of resonant excitation, i.e., $\omega=\omega_0$, the populations given by Eq.~(\ref{rabi}) as a function of time are plotted in Fig.~\ref{fig:rabitrans} (for $F_{\rm DC}=22.2\ \mbox{kV/m}$, $F_{\rm AC}=44.5\ \mbox{kV/m}$), and match well with the numerical solutions obtained from the three-state calculation.

Complete transfer is achieved by applying a $\pi$-pulse, and the time for this is
\begin{equation}
\label{rabitime}
t_{\pi}=\frac{\pi\hbar}{|V|}=\frac{1}{F_{\rm AC}}\frac{\pi\hbar}{q_e\bra{LL}(x_{1}+x_{2}-R)\ket{LR}}=\frac{M}{F_{\rm AC}},
\end{equation}
where
\begin{equation}
M=\frac{\pi\hbar}{q_e\bra{LL}(x_{1}+x_{2}-R)\ket{LR}}.
\end{equation}
The transition time is inversely proportional to the field strength $F_{\rm AC}$.  To a very good approximation, the time given by the analytic Rabi solution is equivalent to the numerical solution which results from the resonant transfer calculations.  Fig.~\ref{fig:rabitrans} shows this and an inset close-up of a selected region of the transition.  The numerical simulation includes all three states, and a small off-resonant coupling to the third state, which is responsible for the oscillations visible in Fig.~\ref{fig:rabitrans}.  We comment on the fidelity of this transfer as a result of these oscillations in Sect.~\ref{Sect:Fidelity}.

\begin{figure}[tb]
\centerline{\includegraphics{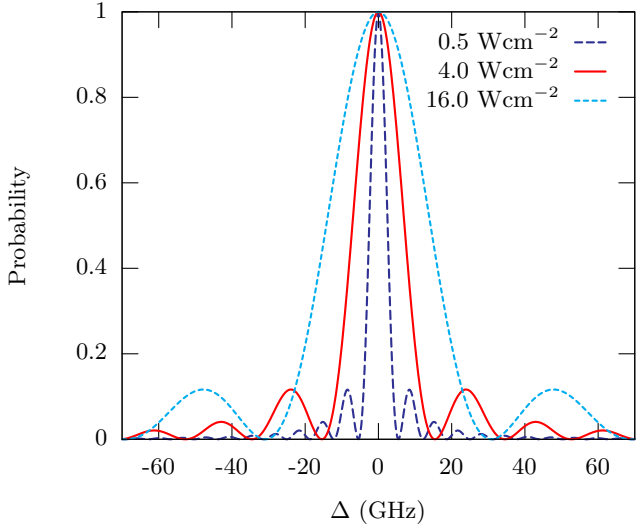}}
\caption{(Colour online) Charge transfer probability as a function of FIR laser detuning for a range of intensities.}
\label{fig:trans_width}
\end{figure}

We also examined the \dzero\dzero$\longrightarrow$\dplus\dmin\ transition probability as a function of detuning, $\Delta$, to observe the response to varying AC field strength.  The results are given in Fig.~\ref{fig:trans_width}, showing the characteristic sinc function dependence and narrowing of the central peak with decreasing field strength.  This well known result serves as a means by which to avoid single donor level transitions.

Spin to charge transduction by gated resonant transfer is a promising technique since the DC selection field is very low compared to the critical field sustainable by the doubly occupied state before electron loss occurs.

\section{Resonant FIR Laser Transfer}
\label{Sect:Optical}
An optically driven version of gated resonant transfer is preferable, given that the separation of the terahertz source and gating circuitry from the rest of the chip, reduces noise from high speed on-chip switching and aids transmission of the signal to the device. An FIR laser operating at wavelengths of $\sim 34\ \mu\mbox{m}$ could provide the required radiation field.  This is on the outer limits of current technology however various candidates exist, including methanol lasers and their deuterated derivatives CD$_3$OH\cite{Sigg}, and possibly synchrotrons and free-electron lasers.  Promising FIR technology also utilises the Si:P system as the active medium for lasing\cite{Pavlov}.  The observation of the \dzero\dzero$\longrightarrow$\dplus\dmin\ transition in optical studies of bulk-doped silicon\cite{Thomas} suggests that this transition may be observed resonantly in this optical version.  

To analyse the optical version, assuming linear polarisation, we rewrite the Hamiltonian of Eq.~(\ref{reshamil})
\begin{equation}
{\cal H}(t)={\cal H}_{0}+{\cal H}_{\rm DC}(t)+{\cal H}_{\rm opt}(t),
\end{equation}
where
\begin{eqnarray}
\label{DChamil}
{\cal H}_{\rm DC}(t)&=&q_e(x_1+x_2-R)F_{\rm DC}(t),\\
\label{opthamil}
{\cal H}_{\rm opt}(t)&=&-\frac{iq_e\hbar}{m}(\vec A\cdot\vec\nabla_1+\vec A\cdot\vec\nabla_2),\\
\label{vecpot}
\vec A(\vec r,t)&=&A_{0}(\omega)\mbox{\boldmath$\hat{\epsilon}$}\big[e^{i(\vec k\cdot\vec r-\omega t)}+e^{-i(\vec k\cdot\vec r-\omega t)}\big].
\end{eqnarray}
Using the dipole approximation, which is valid for the wavelength of the FIR field required here (since $\vec k\cdot\vec R \ll 1$), the Hamiltonian matrix elements for the perturbation reduce to equivalent form
\begin{multline}
4q_e^2\omega^{2}_{21}|A_{0}(\omega)|^2\cos^2\omega t|\bra{\Psi_{LL}}\hat\epsilon\cdot \hat x\ket{\Psi_{LR}}|^2=\\
q_e^2F_{\rm AC}^{2}\sin^2\omega t|\bra{\Psi_{LL}}(x_1+x_2-R)\ket{\Psi_{LR}}|^2.
\label{matrixelements}
\end{multline}
This yields the following relationship for the amplitudes of the vector potential and AC field at resonance,
\begin{equation}
|A_{0}(\omega_0)|^{2}=\frac{F_{\rm AC}^2}{4\omega_0^2}=\frac{M^2}{4\omega_0^2 t_\pi^2},
\end{equation}
and the previous analysis can be applied. Thus, resonant transfer can in principle be achieved via FIR laser excitation.  To do this, the frequency of the laser should be set to the energy gap between the \ket{LR} and \ket{LL} states.  We simulate the transition using the hydrogenic wave functions described in Eqs.~(\ref{phiLR})-(\ref{phiRR}).

It is essential that the small DC offset $F_{\rm DC}$, which serves the purpose of qubit selection, is smaller than the critical DC field strength, $F_{\rm DC}^*$ as outlined in Sect.~\ref{Sect:GRST}.  Staying below this critical value will ensure that ionisation of the \dmin\ state to the conduction band does not occur.  We show the energy levels as a function of the DC field strength, $F_{\rm DC}$, for a donor separation of $R=30\ \mbox{nm}$ in Fig.~\ref{fig:elevels30nm}.
\begin{figure}[tb]
\centerline{\includegraphics{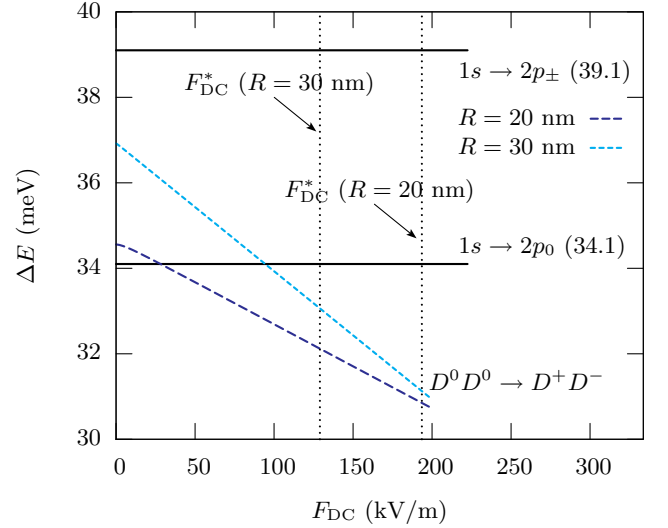}}
\caption{(Colour online) Scaling of the energy gap for the \dzero\dzero$\rightarrow$\dplus\dmin\ (\ket{LR}$\rightarrow$\ket{LL}) transition as a function of $F_{\rm DC}$.  Simulation results are shown for donor separations of $R=20\ \mbox{nm}$ and $R=30\ \mbox{nm}$ against the allowed single donor energy levels (relative to the donor ground state).}
\label{fig:egap}
\end{figure}

It is also important to examine the scaling of the energy gap between states \ket{LR} and \ket{LL} as a function of the DC offset $F_{\rm DC}$.  We do this for donor separations of $R=20\ \mbox{nm}$ and $R=30\ \mbox{nm}$. Fig.~\ref{fig:egap} shows the results against the relevant single donor levels (relative to the ground state).  This will ensure that $F_{\rm DC}$ may be chosen to avoid exciting these single donor transitions.  We also note that there will be no linear Stark effect for the relevant single donor levels\cite{Kohn,Smit} and hence these energy levels will remain unperturbed to first order.  The results are shown in Fig.~\ref{fig:egap} for the dipole allowed transitions, $1s\rightarrow 2p_0$ and $1s\rightarrow 2p_\pm$.  Spectroscopic observations in bulk doped silicon\cite{Thomas}  show that the widths of the $1s\rightarrow 2p$ transitions are considerably less than $1\ \mbox{meV}$, and can be neglected compared to the \dzero\dzero$\longrightarrow$\dplus\dmin\ power broadened transition width.  We note that keeping the laser intensities low will also reduce the probability of causing off-resonant transitions as explained in Sect.~\ref{Sect:GRST}.  The qubit selection field, $F_{\rm DC}$, is chosen to be below $F_{\rm DC}^*$ ($\sim 130\ \mbox{kV/m}$ for $R=30\ \mbox{nm}$) and to avoid single donor transitions.

The time scale of the FIR induced transfer is controlled directly by the laser intensity.  The required laser intensity is
\begin{equation}
I(\omega_0)=\frac{1}{2}\varepsilon_0c\omega_0^2|A_0(\omega_0)|^2=\frac{1}{8}\varepsilon_0c\frac{M^2}{t_\pi^2}.
\label{radfieldint}
\end{equation}
For a charge transfer time of order nanoseconds the required laser power is of order a few milliwatts.  This is in the regime of fast transfer given that $T_{\rm SET}\approx 1\ \mu\mbox{s}$\cite{Buehler}.  At the same time the required laser wavelength can be varied by altering the strength of the local DC field, $F_{\rm DC}$, allowing some flexibility in the requirement for a $34\ \mu\mbox{m}$ FIR laser.  Restrictions on the value of $F_{\rm DC}$ (as discussed earlier) are required in order to avoid coupling to single donor transitions or causing electron loss.

\begin{figure}[tb]
\centerline{\subfigure{\label{fig:opticaltrans20nm}\includegraphics{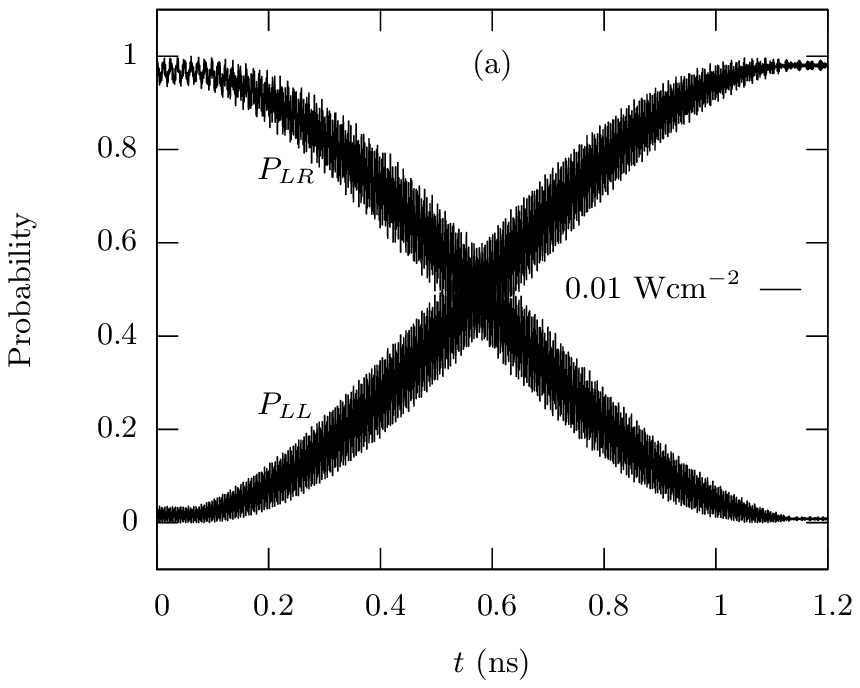}}}
\centerline{\subfigure{\label{fig:opticaltrans30nm}\includegraphics{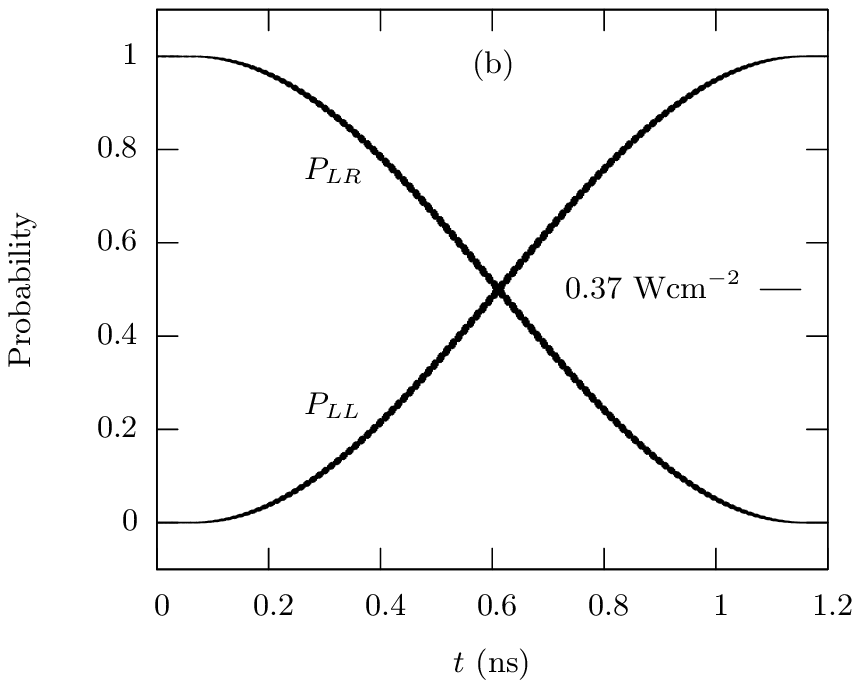}}}
\caption{State probabilities for the resonant FIR laser transfer between donors.  Simulation results are shown for donor separations of (a) $R=20\ \mbox{nm}$ and (b) $R=30\ \mbox{nm}$.}
\label{fig:opticaltrans}
\end{figure}

For donor separations less than $R=30\ \mbox{nm}$ there is larger coupling to the off-resonant state, $\ket{RR}$, due to a larger dipole matrix element (see Fig.~\ref{fig:opticaltrans20nm}).  An example of the low transfer fidelity of such a transition is seen in Fig.~\ref{fig:fidelity20nm}.  Increasing the detuning from this off-resonant state will in principle improve the fidelity, however the local DC field must remain below the critical DC field strength, $F_{\rm DC}^*$, which limits the process.  This suggests that donor separations less than $30\ \mbox{nm}$ will not be practical.  In Fig.~\ref{fig:opticaltrans} we give examples of FIR transfer for separations of $R=20\ \mbox{nm}$ and $R=30\ \mbox{nm}$.  The reduction in the off-resonant coupling with increased donor separation is prominent.

\section{Charge Transfer Fidelity}
\label{Sect:Fidelity}
\begin{figure}[tb]
\centerline{\includegraphics{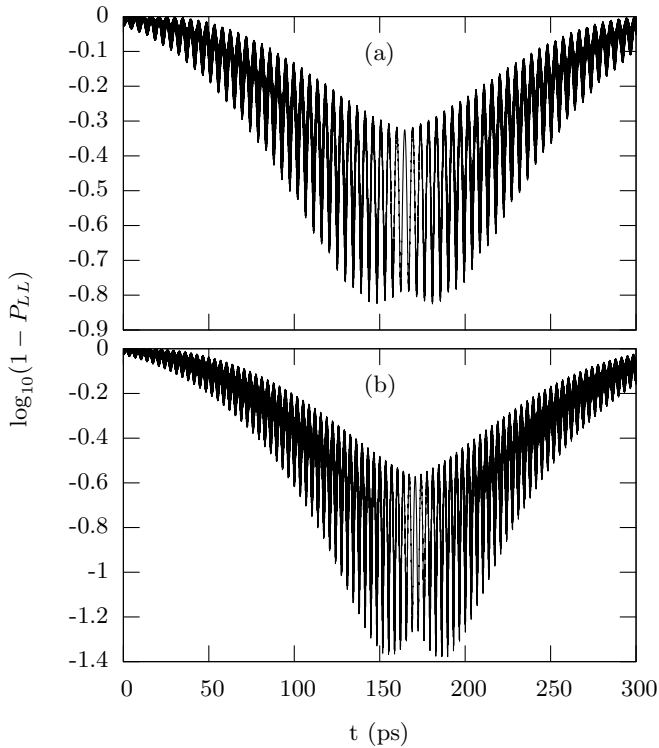}}
\caption{Fidelity of transfer for an FIR laser intensity of $0.37\ \mbox{Wcm}^{-2}$ and donor separation of $R=20\ \mbox{nm}$.  The improved fidelity is noticeable with increased detuning.  Above we show results for $F_{\rm DC}$ values of (a) $8.9\ \mbox{kV/m}$ and (b) $22.2\ \mbox{kV/m}$.  Note that the scales on the y-axes are different.}
\label{fig:fidelity20nm}
\end{figure}

Fidelity of charge transfer is dependent both upon donor separation and detuning from the off-resonant state.  For a given separation, fidelity may be improved by increasing the detuning which is achieved by increasing the local DC field, $F_{\rm DC}$ (see Fig.~\ref{fig:fidelity20nm}).  This process is of course limited by the critical DC field strength, $F_{\rm DC}^*$ and the effect is small.

\begin{figure}[tb]
\centerline{\includegraphics{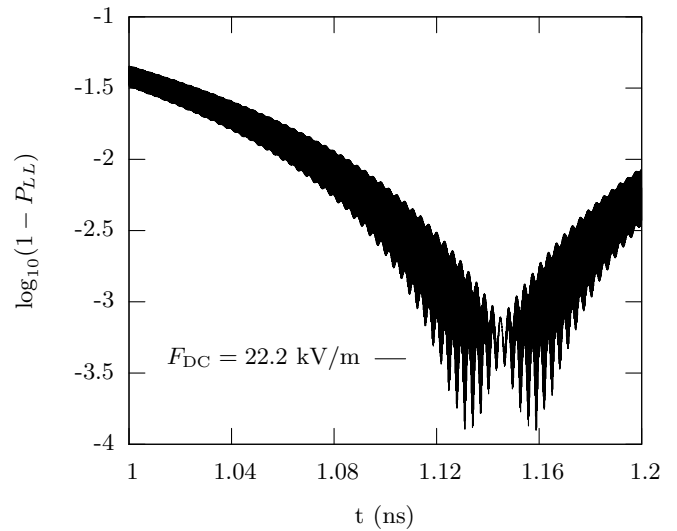}}
\caption{Increasing the donor separation to $R=30\ \mbox{nm}$ reduces the off-resonant dipole matrix element, resulting in improved fidelity (as compared to a donor seaparation of $R=20\ \mbox{nm}$, seen in Fig.~\ref{fig:fidelity20nm}).  Transfer fidelity is shown for an FIR laser intensity of $0.37\ \mbox{Wcm}^{-2}$.}
\label{fig:fidelity30nm}
\end{figure}

Increasing the donor separation in turn reduces the dipole matrix elements to unwanted states which results in smaller off-resonant oscillations, thereby improving fidelity (see Figs.~\ref{fig:fidelity20nm} and \ref{fig:fidelity30nm}).

\begin{figure}[tb]
\centerline{\includegraphics{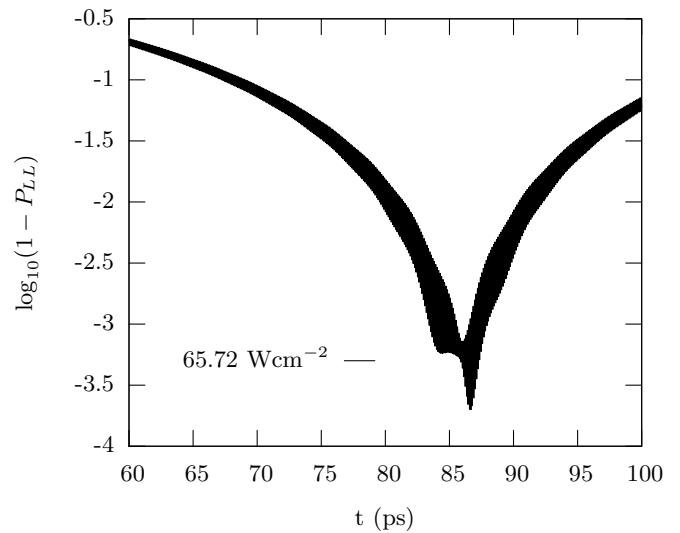}}
\caption{Simulation results for a donor separation of $R=30\ \mbox{nm}$ showing fast, high-fidelity transfer.  Transfer on these time scales requires higher powered lasers which may cause heating and unwanted single donor transitions.}
\label{fig:sharp_fidelity}
\end{figure}

Maximising fidelity should be achieved by careful selection of $F_{\rm DC}$ as well as an understanding of the timing window for FIR laser pulsing.  Low powered lasers are preferential to avoid heating, minimise unwanted single donor transitions, and increase the timing window over which high-fidelity transfer may occur.  Faster transfer is possible, however the timing window over which the FIR laser must be pulsed to achieve high-fidelity transfer is narrow (see Fig~\ref{fig:sharp_fidelity}).  Such high speed switching is possible using laser activated semiconductor switches\cite{Doty,Hegmann}.

Within the approximations used in this preliminary analysis we find greater than $99\%$ transfer fidelity with ample scope for improvement.  Read-out need not operate at the $10^{-4}$ error threshold demanded of logic gates, provided that logic gates can be operated at this threshold or better\cite{Fowler}.

\section{Singlet-Triplet Read-out for Electron Spin Qubits}
\begin{figure}[tb]
\includegraphics{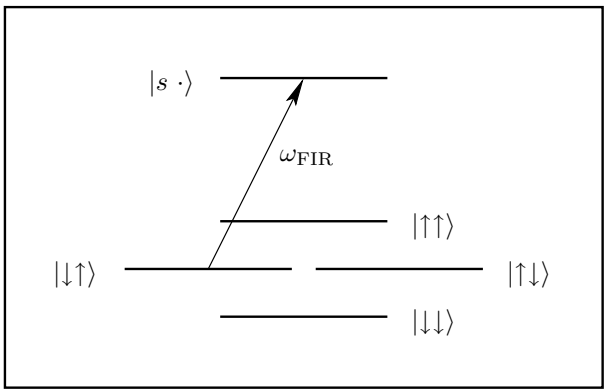}
\caption{Electron spin energy levels for the spatially separated and single site states.}
\label{fig:electron_spin_energy_levels}
\end{figure}

Single-spin read-out fails for the Si:P electron spin SSQC when the $1s$ energy levels are split by an externally applied magnetic field, $B$, as in Fig~\ref{fig:electron_spin_energy_levels}.    The convention used here labels the state of each site, such that \ket{s\ \cdot} represents the doubly occupied spin singlet state formed on the left donor.  Read-out fails in this paradigm given that the states \ket{\negthickspace\downarrow\uparrow}, \ket{\negthickspace\uparrow\downarrow} are degenerate and have equal dipole matrix elements for the \dzero\dzero$\longrightarrow$\dplus\dmin\ transition.  This means the \ket{\negthickspace\downarrow\uparrow} state cannot be preferentially selected without first lifting the degeneracy.

Spectrally resolving the \ket{\negthickspace\downarrow\uparrow} and \ket{\negthickspace\uparrow\downarrow} would provide a physically interesting system to study direct charge transfer and re-initialisation:  an inhomogeneous magnetic field provides one such mechanism for this.  Abe \emph{et~al.}\cite{Abe} proposed a SSQC architecture that relies on a dysprosium (Dy) micromagnet to generate a gradient magnetic field in order to selectively access different nuclear spin qubits using a resonant field.  These Dy micromagnets can generate field gradients of order $20\ \mbox{T}/\mu \mbox{m}$\cite{Goldman}, resulting in a state separation of $7\times 10^{-2}\ \mbox{meV}$  for $R=30\ \mbox{nm}$.  This yields better than $99.99\%$ transfer fidelity.  If instead we apply a site selective hyperfine interaction, $A$, above one donor and not the other in the Si:P electron spin SSQC, the resultant state separation is only $2A\approx 2.4\times 10^{-4}\ \mbox{meV}$.  This is not sufficiently resolved to perform single spin read-out and re-initialisation and only $50\%$ transfer fidelity is achieved.

In a 2-D SSQC architecture with separated read-out and interaction zones, spin resolved read-out and initialisation may be possible.  Such separation requires qubit transfer, and examples include the shuttling process of Skinner \emph{et~al.}\cite{Skinner} and adiabatic passage\cite{Hollenberg2,Greentree2}.  By performing read-out away from the qubit interaction zone,  a Dy micromagnet could be included to provide the required frequency resolution as described above.  Having read-out off site means that the field gradient from the Dy micromagnet would in principle only affect read-out qubits.  Thus qubits in the interaction zone would not need to be characterised to allow for the gradient.  Alternatively, one may use localised magnetic fields (similar to those produced by the current-carrying wire array structures of Lidar \emph{et~al.}\cite{Lidar}).

Performing read-out in the Si:P electron spin quantum computer with the states \ket{\negthickspace\downarrow\uparrow}, \ket{\negthickspace\uparrow\downarrow} degenerate (see Fig.~\ref{fig:electron_spin_energy_levels}) results in singlet-triplet read-out.  Such a scheme could be used for cluster state computation\cite{Rudolph}.  Tuning the FIR laser to the energy difference between the \ket{s\ \cdot} and the degenerate \ket{\negthickspace\downarrow\uparrow}, \ket{\negthickspace\uparrow\downarrow} states results in charge transfer provided the spatially separated states are in an anti-symmetric (singlet) superposition.  The absence of charge detection by the SET projects the electrons onto the spin triplet manifold after read-out.

\section{Conclusion}
Optically driven single-spin read-out for the nuclear spin SSQC via use of an FIR laser has been investigated as an alternative to the adiabatic transfer method of Kane\cite{Kane}.  High-fidelity transfer was shown to be possible on pico to nanosecond time scales which is fast compared to the time required for high-fidelity single-shot measurement by an rf-SET.  We explain how singlet-triplet read-out can be performed in an electron spin paradigm and suggest that it may be used for measurement in cluster state quantum computation.  Spectral resolution of the degenerate states in the electron spin SSQC architecture allows direct single-spin read-out and re-initialisation.  We note that the methods developed are in principle adaptable to any buried donor system.

\section*{Acknowledgments}
LCLH acknowledges discussions with M. Friesen.  The authors also thank A. G. Fowler for his valuable assistance.  This work was supported by the Australian Research Council, the Australian government and by the US National Security Agency (NSA), Advanced Research and Development Activity (ARDA) and the US Army Research Office (ARO) under contract number W911NF-04-1-0290.


\end{document}